\documentclass[12pt]{IOPART}
\usepackage{graphicx}

\def\hc2{$H_{c2}$}

\def\cuscn{$\kappa$-(BEDT-TTF)$_2$Cu(NCS)$_2$}

\def\nh4{$\alpha$-(BEDT-TTF)$_2$NH$_4$Hg(SCN)$_4$}

\newcommand{\ltsim}{\mbox{{\raisebox{-0.4ex}{$\stackrel{<}{{\scriptstyle\sim}}
$}}}}

\begin{document}

\title[Deuteration of \cuscn]
{Comparison of the normal state properties of
\cuscn ~and its deuterated
analogue in high
magnetic fields and under high hydrostatic pressures}
\author{Tim Biggs$^1$, Anne-Katrin Klehe$^1$,
John Singleton$^{1,2}$,\footnote[3]{To
whom correspondence should be addressed (j.singleton1@physics.ox.ac.uk)}
David Bakker$^1$, Jane Symington$^1$,
Paul Goddard$^1$, Arzhang Ardavan$^1$,
William Hayes$^1$
John A~Schlueter$^3$, Takehiko Sasaki$^4$ and Mohamedally Kurmoo$^5$}

\address{$^1$~Department of Physics, University of Oxford, The
Clarendon Laboratory, Parks Road, Oxford OX1 3PU, United Kingdom}

\address{$^2$~National High Magnetic Field Laboratory, LANL, MS-E536, Los
Alamos, New Mexico 87545, USA}

\address{$^3$~Materials Science Division,
Argonne National Laboratory, Illinois 60439, USA}

\address{$^4$~Institute for Materials Research, Tohoku University,
Aoba Ku, Sendai, Miyagi~9808577, Japan}

\address{$^5$~IPCMS, 23 rue du Loess, BP~20/CR, 67037~Strasbourg~Cedex,
France}

\begin{abstract}
{\sloppy
Details of the Fermi-surface topology
of deuterated \cuscn ~have been measured
as a function of pressure, and compared
with equivalent measurements of the undeuterated salt.
We find that the superconducting transition temperature is
much more dramatically suppressed by increasing pressure in the
deuterated salt. It is suggested that this is linked to
pressure-induced changes
in the Fermi-surface topology, which occur
more rapidly in the deuterated salt than in the undeuterated
salt as the pressure is raised.
Our data suggest that the negative isotope effect
observed on deuteration is due to small differences
in Fermi-surface topology caused by the isotopic substitution.
}
\end{abstract}

\submitto{\JPCM}

\maketitle

The nature of superconductivity in quasi-two
dimensional crystalline organic metals is the subject of current
debate in the literature~\cite{john2,john,schmalian,kuroki}. The
close proximity of an antiferromagnetic to a superconducting
groundstate in the temperature-pressure phase diagram has spurred
theoretical suggestions of d-wave Cooper pairing mediated by
antiferromagnetic fluctuations~\cite{schmalian,kuroki,charffi,aoki,lee}. Such
an idea, which implies nodes in the superconducting order
parameter, is strongly supported by NMR ($^{13}$C~\cite{NMR,NMRa,NMRb}
and $^1$H~\cite{NMR2}), tunnelling~\cite{STM}, thermal
conductivity~\cite{sasaki} and magnetic penetration depth
experiments~\cite{carrington}, and by the form of the superconducting
phase diagram deduced from magnetometry and
NMR~\cite{french}. The coupling of
Raman modes to the antiferromagnetic fluctuations
has also been observed~\cite{eld1}, suggesting interactions between the
lattice and the magnetic fluctuations. On the other hand,
it has been suggested that
specific heat measurements may be interpreted using
a BCS-like model~\cite{wosnitza}. The observed
hardening of low energy, intramolecular vibrations at
the superconducting transition in Raman~\cite{raman,photonself} and
inelastic neutron scattering
experiments~\cite{pintschovius} has been interpreted as further
evidence for the involvement of phonons in superconductivity.
However, similar phonon self-energy effects have also been
observed in the non-BCS like Cuprate superconductors~\cite{Y123}
and perhaps merely indicate very strong
electron-phonon coupling.

In this context, the observation of a
``negative isotope effect'' in
\cuscn ~may be of great importance~\cite{kini,schlueter};
on replacing the
terminal hydrogens of the BEDT-TTF molecule
in \cuscn ~by deuterium, it was found that
a small but consistent increase in the
superconducting critical temperature
$T_{\rm c}$ occurred~\cite{kini,schlueter}.
By contrast, isotopic substitutions of other heavier
atoms in the BEDT-TTF molecule or in the anion layer exhibit a
very small, normal isotope effect or no significant isotope
effect at all, respectively~\cite{schlueter}.
We have therefore studied the changes
in Fermi-surface parameters of \cuscn ~on deuteration
using the Shubnikov-de Haas effect.
Data were recorded both at ambient pressure and as
a function of hydrostatic pressure.
Taken in conjunction with very recent millimetre-wave
magnetoconductivity experiments~\cite{rachel}, our data suggest that
it is primarily the changes in the detailed topology of the
of the Fermi surface
brought about by deuteration that cause the observed
isotope effect.
This would tend to support models for superconductivity
involving pairing via electron-electron
interactions~\cite{schmalian,kuroki,charffi,aoki,lee}.

The experiments involved single crystals of
\cuscn  ~($\sim 0.7 \times 0.5 \times 0.1$~mm$^3$; mosaic spread
$\ltsim 0.1^{\circ}$), produced using
electrocrystallization~\cite{kini,schlueter}.
In some of the crystals, the terminal hydrogens of the BEDT-TTF molecules
were isotopically substituted by deuterium;
we refer to these deuterated samples as d8, and
conventional hydrogenated samples
as h8. In order to check for extrinsic effects,
independently-prepared batches of both
types of crystal were made at Argonne, Strasbourg and
Sendai; no extrinsic effects were found.
Note that, to all intents and purposes, the crystallographic
unit cells of h8 and d8 seem to be indistinguishable
in size and shape~\cite{schlueter,watanabe,perscom}.

The magnetoresistance of the samples was measured using
standard 4 wire AC-techniques (frequency $f=15-180$~Hz,
current $I=1-20~{\mu}$A)~\cite{john}.
Contacts were applied to the upper and lower large surfaces
of the crystals, so that the current was directed
and the voltage measured in the
interlayer direction; such a configuration
gives a resistance which is accurately proportional to
the interlayer component of the magnetoresistance, $\rho_{zz}$~\cite{john}.

In the ambient-pressure experiments, crystals
of d8 and h8 were simultaneously studied in a $^3$He/$^4$He cryostat
which allowed rotation of the samples to all possible angles in
the magnetic field~\cite{goddard}. The magnetoresistance was measured
with the samples at many orientations in the magnetic field,
so that any errors due to slight differences of mounting of the
d8 and h8 crystals could be eliminated;
the quasiparticle effective masses and
Shubnikov-de Haas oscillation frequencies
discussed below are corrected to $\theta =0$,
where $\theta$ is the angle between the
normal to the sample's conducting planes and the
magnetic field.
Temperatures were monitored using ruthenium oxide sensors,
with additional checks carried out using the $^3$He
and $^4$He vapour pressures.
Quasistatic magnetic fields were provided
by a 15~T superconductive magnet at Los Alamos
and by the 45~T Hybrid magnet at NHMFL Tallahassee.

The high-pressure experiments were carried out
on three d8 crystals
using a non-magnetic piston-cylinder cell; the pressure medium
was Fluorinert FC75~\cite{eremets}.
The exact crystal orientation with respect to the field was determined by
comparison with ambient pressure data and corrections
to the measured Shubnikov-de Haas frequency
made accordingly~\cite{john}.
The cell was placed in a large volume $^3$He
cryostat capable of temperatures down to 700~mK
within a 17~T superconductive magnet at Oxford.
The pressure inside the cell was determined using a manganin
wire~\cite{eremets}. Temperatures were
measured with a Pt thermometer above 50~K and a ruthenium oxide
thermometer at low temperatures. All pressures quoted were measured
at 4.2~K.
$T_{\rm c}$ was taken to be the resistive
midpoint of the normal-superconducting transition during cool
down.

\begin{figure}[htbp]
\centering
\includegraphics[height=8cm]{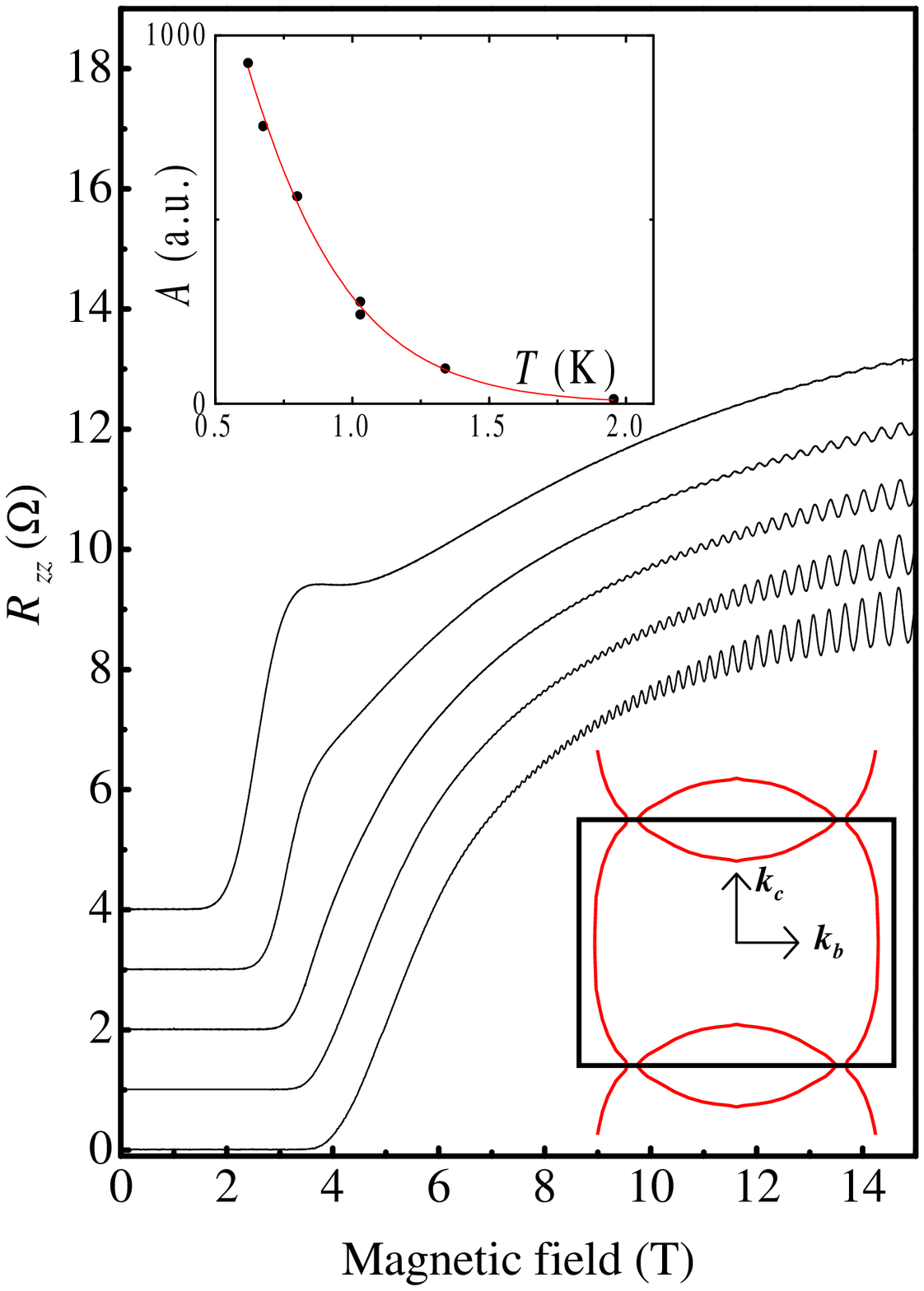}
\includegraphics[height=8cm]{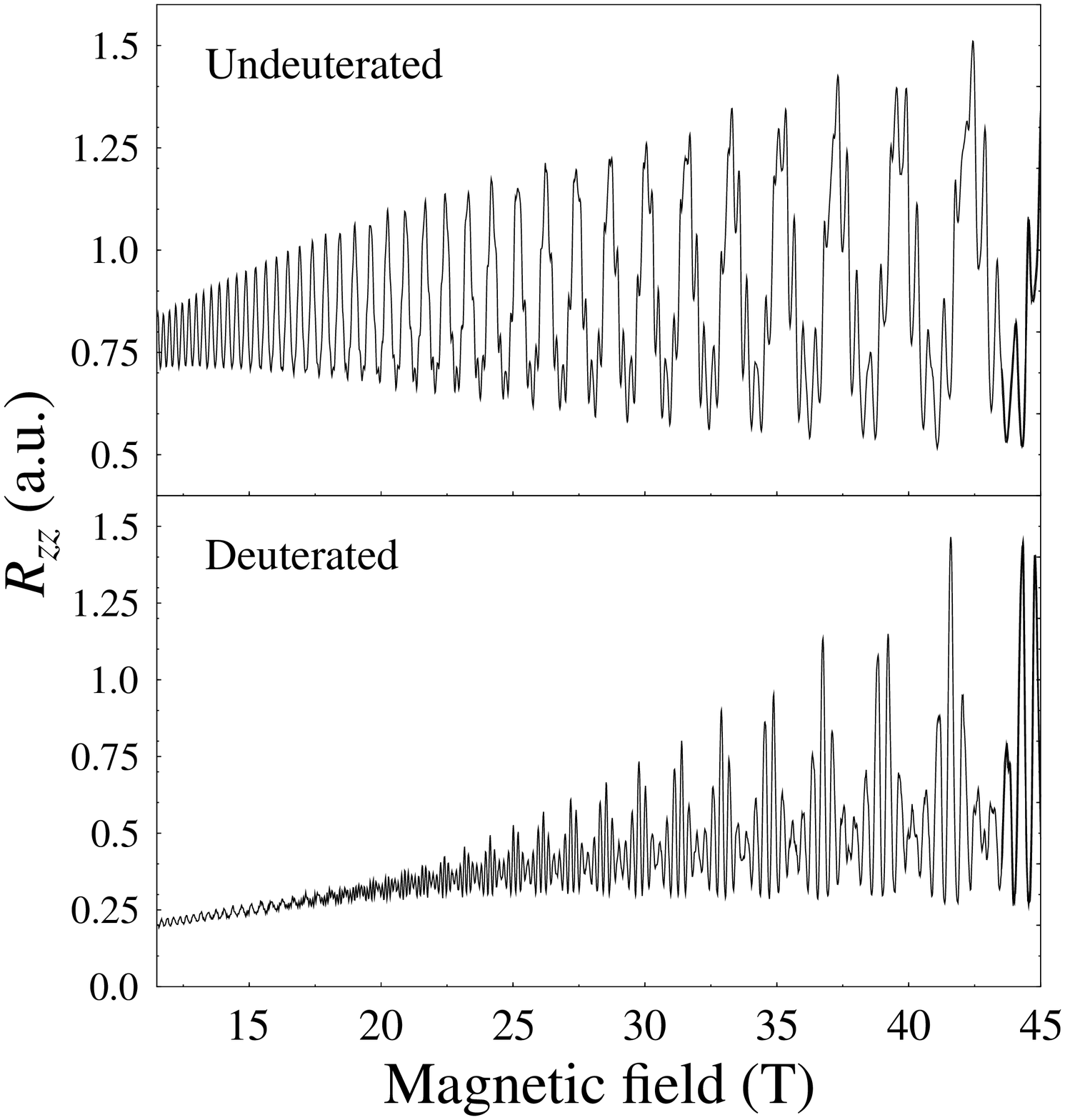}
\caption{Left: interplane resistance $R_{zz}$ ($\propto \rho_{zz}$)
of h8 \cuscn ~at ambient pressure with
magnetic field applied perpendicular to the quasi-two-dimensional planes.
Data for temperatures 1.96~K (uppermost trace),
1.34~K, 1.03~K, 800~mK and 620~mK (lowest trace)
are shown; for clarity, the data have been offset by
$1~\Omega$.
The superconducting to normal transition is clearly
visible, as are Shubnikov-de Haas oscillations due to
the $\alpha$ pocket of the Fermi surface.
The upper inset shows a typical
plot of the Fourier amplitude $A$
of the Shubnikov-de Haas oscillations
as a function of temperature; data are points
and the curve is a fit of the
Lifshitz-Kosevich formula~\cite{john}.
The lower inset shows the Brillouin zone
and Fermi surface cross-section of \cuscn ~with its
closed
$\alpha$ pocket and quasi-one-dimensional
sheets (based
on parameters given in
Reference~\cite{goddard}).
Right: comparison of the high-field interplane resistance $R_{zz}$
of d8 (deuterated) and h8 (undeuterated)
\cuscn ~($T=520$~mK).
Note how the high frequencies caused by magnetic breakdown are
much more dominant in d8.
}
\label{fig1}
\end{figure}

The left-hand side of
Figure~\ref{fig1} shows typical low-field ambient-pressure
magnetoresistance data for an h8 sample.
Similar data were recorded simultaneously
for a d8 sample.
Shubnikov-de Haas oscillations caused by the
quasi-two-dimensional $\alpha$ pocket of
the Fermi surface (see lower inset of Figure~\ref{fig1}).
are visible.
At these low fields, the oscillatory
magnetoresistance is much less than
the non-oscillatory component, and magnetic breakdown
is a relatively minor consideration~\cite{john}.
Hence, the Lifshitz-Kosevich
formula may be used to extract
the effective mass $m^*$~\cite{john}
from the temperature dependence of the
oscillation amplitude $A$;
a typical fit is shown as the upper inset
in Figure~\ref{fig1}.

Data such as those in Figure~\ref{fig1}
suggest that the $\alpha$ Fermi-surface pockets of
h8 and d8 are rather similar; as an example,
the magnetic quantum oscillation frequencies of the
$\alpha$ pocket at $\theta=0$
were $F_{\alpha} = 600 \pm 1$~T (h8) and
$F_{\alpha} = 597 \pm 1$~T (d8); the former is in good agreement
with the accepted value~\cite{john2}. Although the d8 and h8
frequencies are very close, consistently smaller
values were obtained for the d8 samples,
and so we believe that the stated difference is real.
The corresponding $\alpha$ pocket effective masses
($\theta =0$)
are $m^* =3.5 \pm 0.1 m_{\rm e}$ (h8) and
$m^* = 3.4 \pm 0.1 m_{\rm e}$ (d8);
the difference between the masses is around
the experimental error.
The average interlayer transfer integrals for d8 and h8 were
measured in a separate experiment~\cite{goddard};
both were found to be very close to 0.04~meV~\cite{goddard}.

The only significant difference between the magnetotransport
of d8 and h8 at ambient pressure
occurs in the magnetic breakdown between the $\alpha$ pocket
and quasi-one-dimensional sheets, which gives rise
to a semiclassical orbit with the same cross-sectional area
as the Brillouin zone~\cite{neilbd}.
As shown in the right-hand side of
Figure~\ref{fig1}, which displays magnetoresistance
data recorded in the hybrid magnet,
the breakdown is significantly stronger in d8,
leading to a plethora of high
frequency oscillations in the magnetoresistance
due to the Shiba-Fukuyama-Stark quantum interference effect~\cite{john}.
Following the method set out in Reference~\cite{neilbd},
analysis of the breakdown oscillations suggests a breakdown field
of $B_0=30 \pm 5$~T in d8, compared to a value of $B_0=41 \pm 5$~T~\cite{neilbd}
in h8.

\begin{figure}[htb]
\centering
\includegraphics[height=8cm]{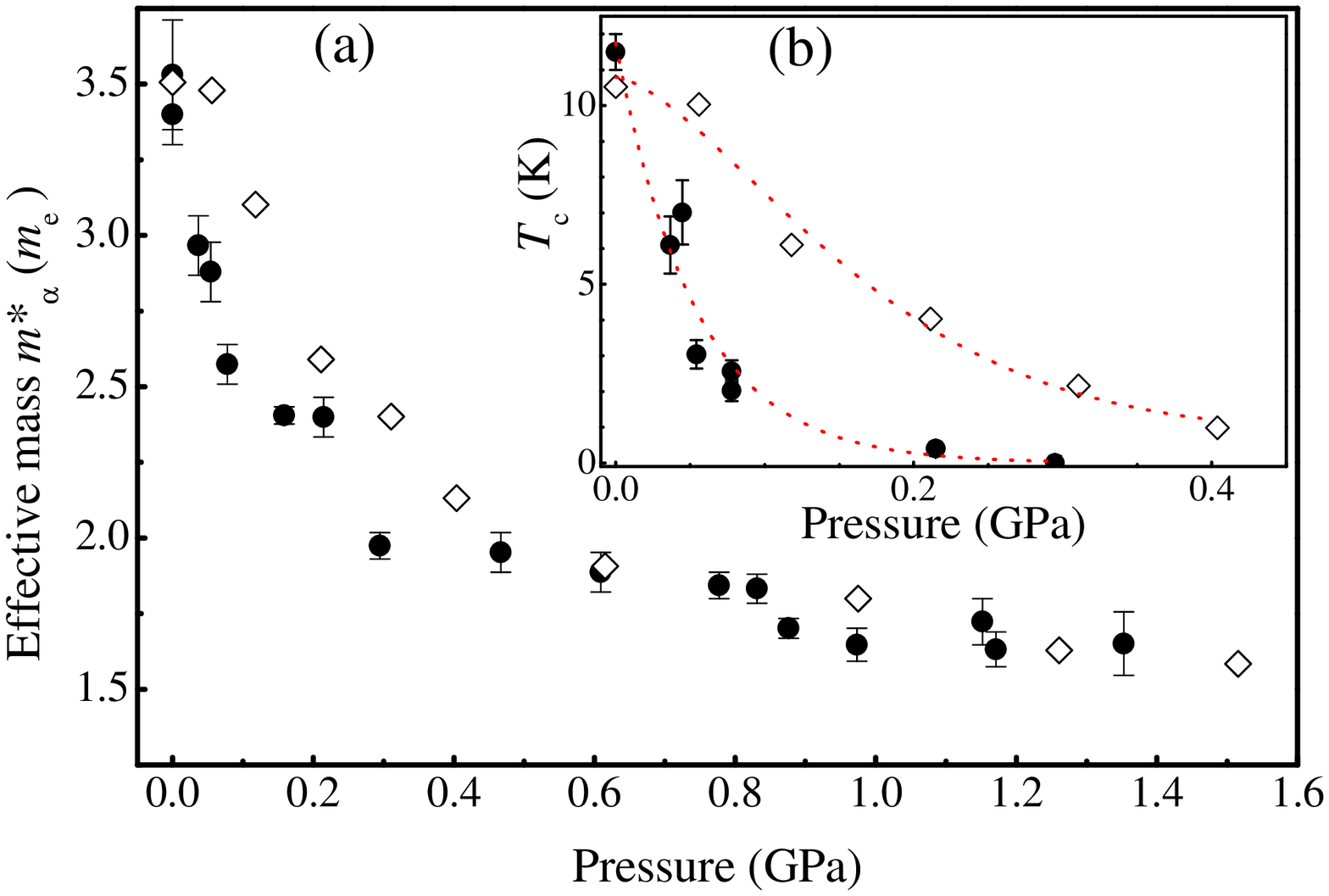}
\includegraphics[height=8cm]{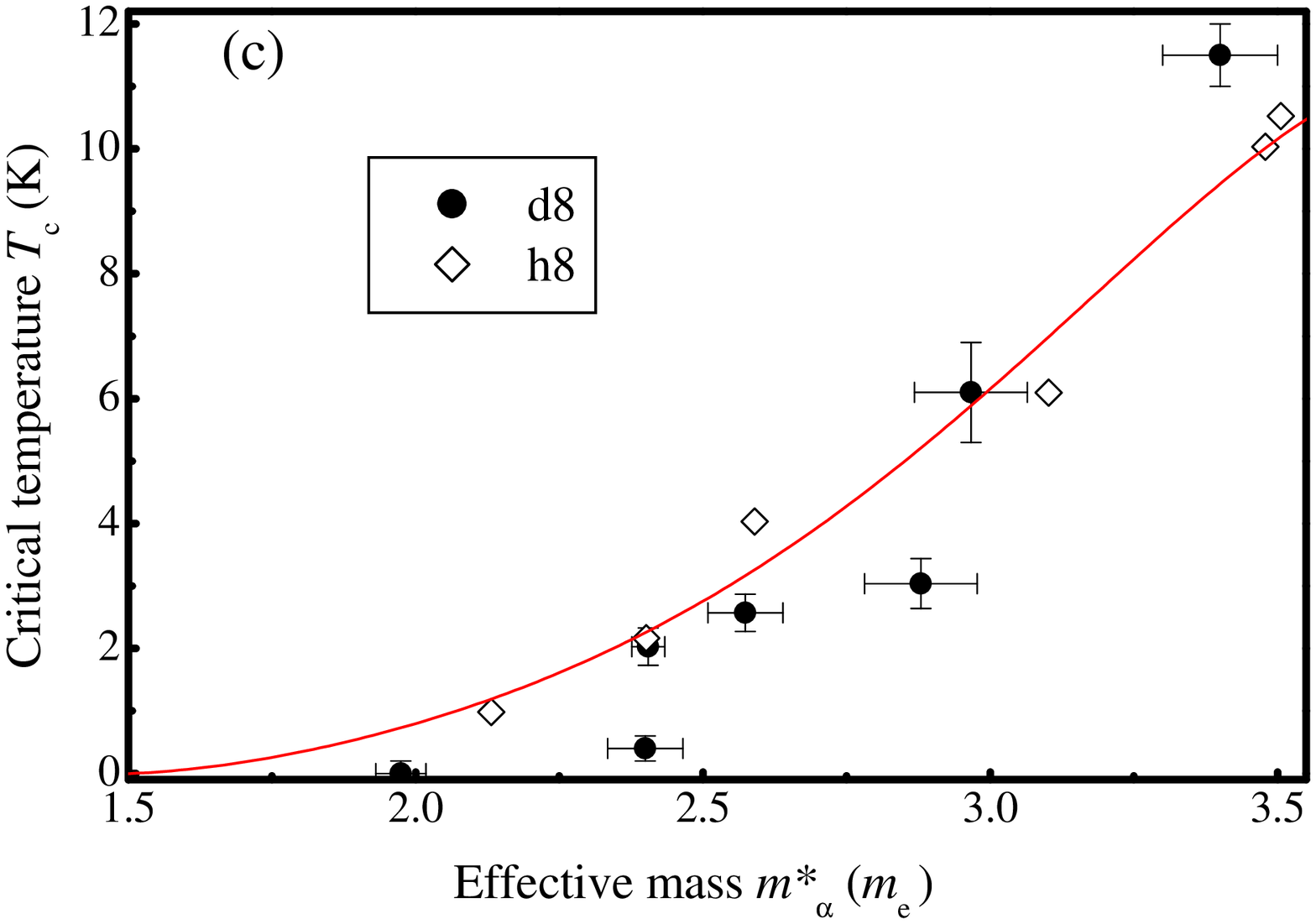}
\caption{(a)~Pressure dependence of the
effective mass $m_{\alpha}^*$
of the $\alpha$ Fermi-surface pocket
of \cuscn
as a function
of pressure $P$. Data for d8 (this work) are filled
circles; data for h8 (Reference~\cite{caulfield})
are hollow diamonds.
(b)~Variation of superconducting critical
temperature $T_{\rm c}$ of \cuscn
as a function of $P$;
filled circles: d8 (this work);
hollow diamonds: h8
(Reference~\cite{caulfield}).
(c)~$T_{\rm c}$ versus $m^*_{\alpha}$
for h8 (filled circles) and d8 (hollow diamonds).
The curve is the linearised Eliashberg solution
from Reference~\cite{lee}.}
\label{fig3}
\end{figure}

We now turn to the high pressure experiments.
Figure~\ref{fig3}(a) shows the
pressure dependence of the effective mass
$m^*_{\alpha}$ of the d8 $\alpha$ Fermi-surface pocket, extracted
from the temperature dependence of the
Shubnikov-de Haas oscillations~\cite{john}.
At pressures of less than 0.25 GPa,
$m_{\alpha}^*$~decreases very sharply with increasing pressure $P$
(${\rm d}m_{\alpha}^*/{\rm d}P \approx -10~m_{\rm e}$/GPa);
above this pressure, the mass decreases much more gently
as $P$ is raised.
The d8 masses are compared with the h8 data of Caulfield
{\it et al.}~\cite{caulfield} in Figure~\ref{fig3}(a);
note that the initial rate of decrease of $m_{\alpha}^*$
with $P$
is significantly less in h8, but that the variation of
$m_{\alpha}^*$ is very similar in h8 and d8 at higher $P$.

Figure~\ref{fig3}(b) shows the
superconducting critical
temperature $T_{\rm c}$ of \cuscn
as a function of $P$ for both d8 (this work)
and h8 (Reference~\cite{caulfield}).
In the case of d8, $T_{\rm c}$ is very
rapidly suppressed with increasing $P$
$({\rm d}T_{\rm c}/{\rm d}P \approx 80$~K/GPa).
By contrast, ${\rm d}T_{\rm c}/{\rm d}P \approx 30$~K/GPa
for h8~\cite{caulfield}. As in the case of the
effective mass, the pressure seems to have a much
more marked effect for d8 that for h8.

Weiss {\it et al.} suggested that there may be some
universal relationship between $T_{\rm c}$
and $m^*_{\alpha}$ in $\kappa$-phase
BEDT-TTF superconductors~\cite{weiss}
(see also References~\cite{caulfield,brooks}).
Figure~\ref{fig3}(c) shows such a plot for d8 (this work)
and h8 (Reference~\cite{caulfield}).
Whilst the data for the two salts vary in a
qualitatively similar fashion,
the slope of $T_{\rm c}$ versus
$m^*_{\alpha}$ seems to be somewhat steeper
for d8.

\begin{figure}[htb]
\centering
\includegraphics[height=8cm]{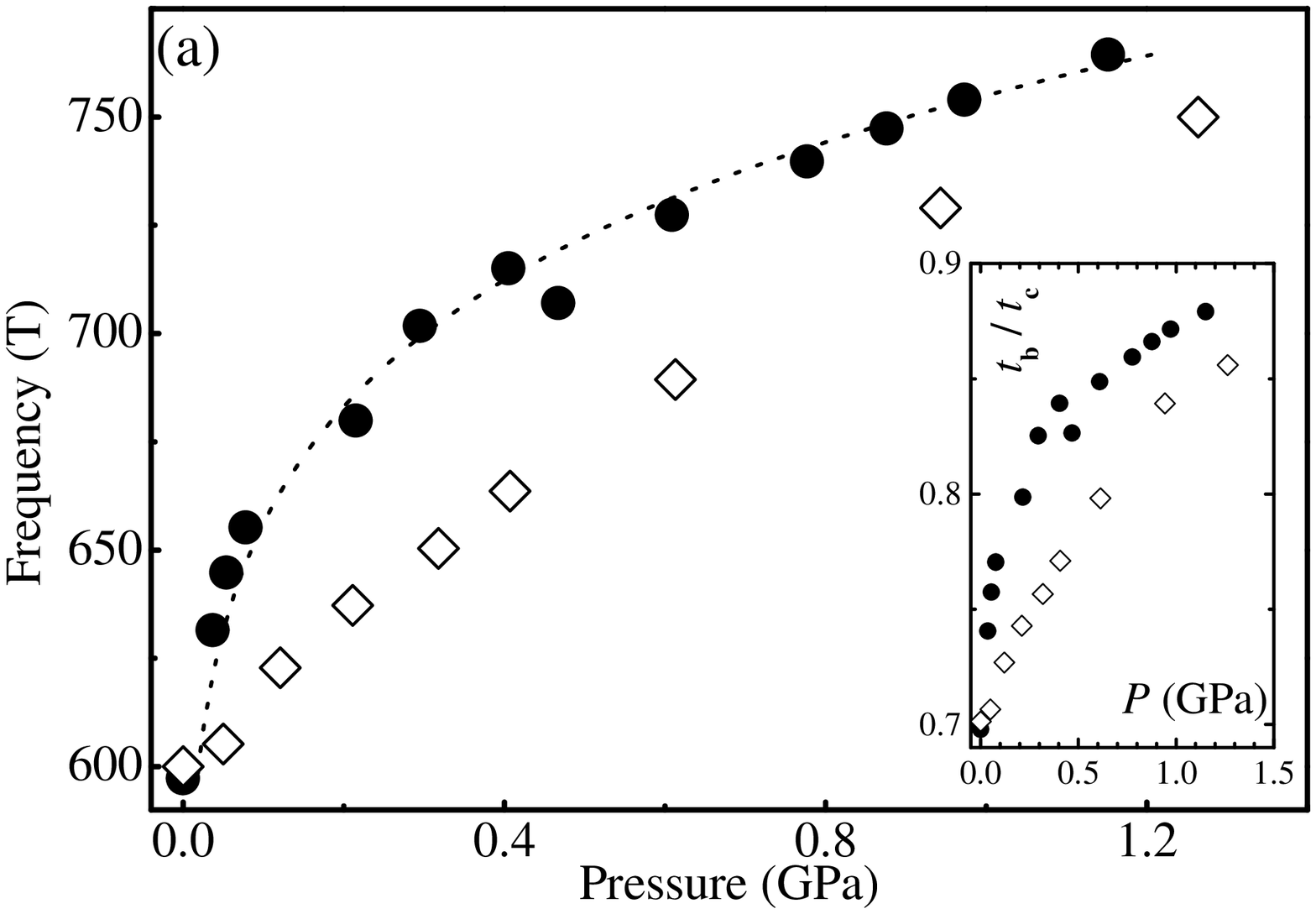}
\includegraphics[height=8cm]{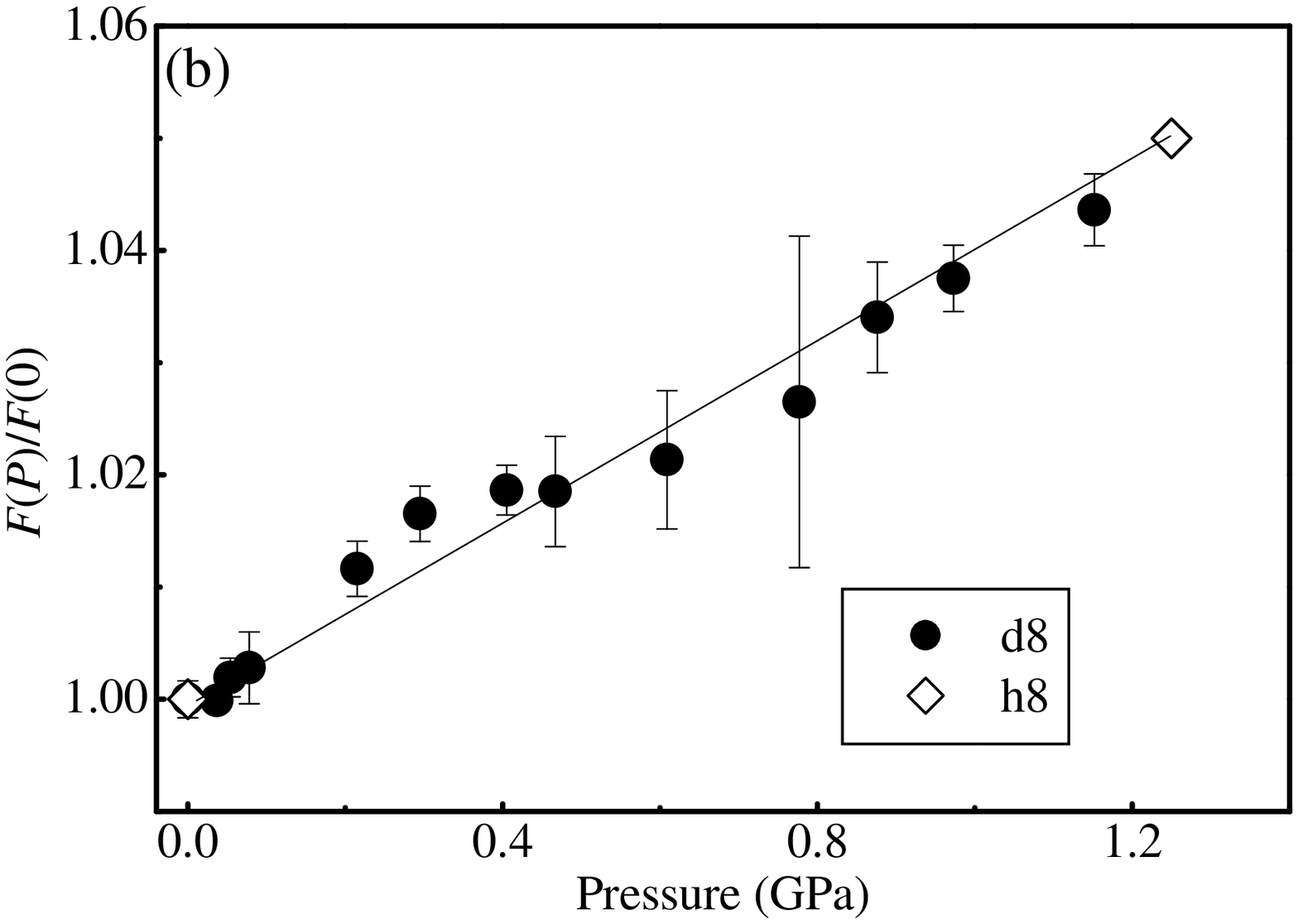}
\caption{(a)~Shubnikov-de Haas oscillation frequency
for the $\alpha$ Fermi-surface pocket as
a function of pressure $P$ for d8 (filled circles; this work)
and h8 (hollow diamonds, Reference~\cite{caulfield}).
The inset shows $t_{\bf b}/t_{\bf c}$ versus pressure,
where the $t$ are effective transfer integrals defined
in Equation~\ref{phwoar}.
(b)~Equivalent plot for the $\beta$ breakdown frequency,
but with the frequencies normalised
to the ambient-pressure value.}
\label{fig4}
\end{figure}

Figures~\ref{fig4}(a) and (b) show the pressure dependence of the
Shubnikov-de Haas oscillation frequencies for d8 and h8.
As has been mentioned above,
the $\beta$-orbit frequency $F_{\beta}$ (Figure~\ref{fig4}(b))
reflects the size of the Brillouin zone in the
conducting {\bf bc} plane; it is therefore a direct measure of the
in-plane compressibility.
Figure~\ref{fig4}(b) shows that the pressure dependence
of $F_{\beta}$ is almost identical in h8 and d8.
This suggests that any lattice
softening effects due to deuteration do not affect the intraplane
compressibility, and are thus, if present, only
effective in the interplane direction.
By contrast, the $\alpha$ Fermi-surface pocket frequency
grows much more quickly with $P$ in d8 than in h8
(Figure~\ref{fig4}(a));
interestingly, $T_{\rm c}$ tends to zero
in {\it both} d8 and h8 at pressures where
the $\alpha$-orbit frequencies
reach approximately the same value, $F_{\alpha} \approx 770 \pm 15$~T.

To examine the effect of pressure more deeply, we turn to the
{\it effective dimer model} which has been shown to
represent the intralayer
quasiparticle dispersion $E({\bf k}_{||})$
in \cuscn ~accurately (see Reference~\cite{goddard}
and references therein);
\begin{equation}
E({\bf k}_{||})=
\pm 2\cos(\frac{k_{\bf b}b}{2})
\sqrt{t_{{\bf c}1}^2+t_{{\bf c}2}^2+2t_{{\bf c}1}t_{{\bf c}2}\cos(k_{\bf
c}c)}
+2t_{\bf b}\cos(k_{\bf b}b).
\label{phwoar}
\end{equation}
Here $k_{\bf b}$ and $k_{\bf c}$ are the intralayer components of {\bf k}
(see Figure~\ref{fig1}, inset) and $t_{\bf b}$, $t_{{\bf c}1}$
and $t_{{\bf c}2}$ are effective interdimer transfer
integrals~\cite{comment};
the $+$ and $-$ signs result in the quasi-one-dimensional
sheets and the $\alpha$ pocket
of the Fermi surface respectively.
The cross-sectional area of the $\alpha$ pocket is
determined by the ratio
$t_{\bf b}/t_{\bf c}$, where $t_{\bf c}$ is the mean
of $t_{{\bf c}1}$ and $t_{{\bf c}2}$~\cite{goddard}.
Using this approach~\cite{caulfield}, we can convert the
frequencies $F_{\alpha}$ from Figure~\ref{fig4}(a) into
values of $t_{\bf b}/t_{\bf c}$; the result is shown
for both h8 and d8 as the inset in Figure~\ref{fig4}(a).
The inset shows that $t_{\bf b}/t_{\bf c}$ increases with
$P$ more rapidly in d8 than in h8.

Reference~\cite{caulfield} shows that an increase of
$t_{\bf b}/t_{\bf c}$ elongates the overall Fermi-surface
cross-section in the $k_{\bf c}$ direction by ``fattening''
the $\alpha$ pocket. As a consequence,
the corrugation of the quasi-one-dimensional sheets changes
somewhat; the regions next to the breakdown gap become slightly
more pointed, whilst away from the gap, the sheets flatten slightly.
Our data suggest that these changes occur much more rapidly with
increasing pressure in d8 than in h8.

With this in mind, we suggest that the more rapid suppression
of superconductivity in d8, compared to h8, is linked to the
fact that the Fermi surface topology changes more
drastically with pressure in d8 (Figure~\ref{fig4}). This strongly
suggests that the superconducting mechanism is very sensitively
influenced by the exact topology of the Fermi surface,
and hence that this effect is also responsible for the inverse isotope
effect in \cuscn ~observed on deuteration.
Additional support for this proposal is provided by
the difference in magnetic
breakdown strength seen in d8 and h8 at ambient pressure
(Figure~\ref{fig1}), suggesting slightly different Fermi-surface
topologies for the two salts.

An alternative explanation for the inverse isotope effect
invokes a softening of the bonds upon
deuteration and a concurrent increase in the electron-phonon
interaction~\cite{schlueter}. Comparative Raman studies on
on deuterated and undeuterated \cuscn ~may also be
interpreted in this way~\cite{pedron2}.
It might also be possible to simulate the size and
direction of change of $T_{\rm c}$ experienced upon deuteration
from the anisotropic compressibility~\cite{watanabe,rahal} of \cuscn,
and the strongly anisotropic uniaxial pressure dependence of its
superconducting transition temperature~\cite{lang,sadewasser,langmoan}.
However, neither of these explanations can shed any light
on the very obvious relationship between
the details of the Fermi-surface shape and $T_{\rm c}$,
shown by our data (Figures \ref{fig3} and \ref{fig4}).

Instead, our data support models for
exotic d-wave superconductivity in the
organics which invoke electron-electron
interactions depending
on the topological (i.e. nesting) properties
of the Fermi surface~\cite{schmalian,kuroki,charffi,aoki}.
Similar interactions probably contribute to the
relatively large values of the quasiparticle mass
observed in \cuscn~\cite{john2}.
Hence, the changes in Fermi-surface topology
may cause {\it both} the suppression
of the superconducting transition temperature
and the effective mass (see Figure~\ref{fig3});
the causal relationship between $T_{\rm c}$ and $m^*$
suggested in earlier
works~\cite{caulfield,weiss,brooks} is perhaps
an oversimplification.

Support for our interpretation comes from recent
millimetre-wave magnetoconductivity experiments which
give information about the corrugations of the
quasi-one-dimensional sheets of the Fermi surface~\cite{rachel}
in the {\it interlayer} direction.
It was found that the corrugations on the Fermi sheets
of h8 (lower $T_{\rm c}$) were relatively large compared
to those on the sheets of d8 (higher $T_{\rm c}$).
This again suggests that it is primarily details
of the Fermi-surface topology, and in particular its
nestability, that determine $T_{\rm c}$.

In summary, we have measured details of the Fermi-surface topology
of deuterated \cuscn ~as a function of pressure, and compared
them with equivalent measurements of the undeuterated salt.
We find that the superconducting transition temperature is
much more dramatically suppressed by increasing pressure in the
deuterated salt. This may be linked to
pressure-induced changes
in the Fermi-surface topology, which occur
more rapidly in the deuterated salt as the pressure is raised.
Our data support models for
exotic d-wave superconductivity in the
organics which invoke electron-electron
interactions depending
on the topological properties
of the Fermi surface.

The work is supported by EPSRC (UK).
We should like to thank Steve Blundell
and Ross McDonald
for stimulating discussions.


\vspace{2cm}
\begin{thebibliography}{99}
\bibitem{john2}
John Singleton and Charles Mielke, Contemp. Phys. {\bf 43}, 150 (2002).
\bibitem{john}
John Singleton, Reports on Progress in Physics, {\bf 63}, 1111
(2000).
\bibitem{schmalian}
J.~Schmalian, Phys. Rev. Lett. {\bf 81} 4232 (1998).
\bibitem{kuroki}
K. Kuroki, T. Kimura, R. Arita, Y. Tanaka and Y. Matsuda,
preprint cond-mat 0108506 (2001).
\bibitem{charffi}
S. Charf fi-Kaddour, A. Ben Ali, M. H\'{e}ritier and R. Bennaceur,
J. Superconductivity {\bf 14}, 317 (2001); R. Louati, S.
Charfi-Kaddour, A. Ben Ali, R. Bennaceur and M. H\'{e}ritier,
Synth. Met. {\bf 103}, 1857 (1999); R. Louati, S. Charfi-Kaddour, A. Ben Ali, R. Bennaceur
and M. Heritier, Phys. Rev. B {\bf 62}, 5957 (2000).
\bibitem{aoki}
K.~Kuroki and H.~Aoki, Phys. Rev. B {\bf 60}, 3060 (1999).
\bibitem{lee}
Won-Min Lee,
Solid State Commun.{\bf 106}, 601 (1998).
\bibitem{NMR}
K.~Kanoda, K. Miyagawa, A. Kawamoto, Y. Nakazawa , Phys. Rev. B
{\bf 54} 76 (1996).
\bibitem{NMRa}
H. Mayaffre, P. Wzietek, D. J\'{e}rome, C. Lenoir and P. Batail ,
Phys. Rev. Lett. {\bf 75} 4122 (1995).
\bibitem{NMRb}
S.M. de Soto, C. P. Slichter ,A. M. Kini, H. H. Wang, U. Geiser
and J. M. Williams , Phys. Rev. B {\bf 54} 16101 (1996).
\bibitem{NMR2}
S. Lefebvre, P. Wzietek, S. Brown, C. Bourbonnais, D. J\'{e}rome,
C. M\'{e}zi\`{e}re, M. Fourmigu\'{e} and P. Batail, Phys. Rev.
Lett. {\bf 85} 5420 (2000).
\bibitem{STM}
T. Arai, K. Ichimura, K. Nomura, S. Takasaki, J. Yamada, S.
Nakatsuji and H. Anzai, Phys. Rev. B {\bf 63} 104518 (2001).
\bibitem{sasaki}
K. Izawa, H. Yamaguchi, T. Sasaki and Y. Matsuda,
Phys. Rev. Lett. {\bf 88}, 027002 (2002).
\bibitem{carrington}
A.~Carrington, I. J. Bonalde, R. Prozorov, R. W. Giannetta, A. M.
Kini, J. Schlueter, H. H. Wang, U. Geiser and J. M. Williams,
Phys. Rev. Lett. {\bf 83} 4172 (1999).
\bibitem{french}
S. Lefebvre, P. Wzietek, S. Brown, C. Bourbonnais, D. Jerome, C. Meziere,
M. Fourmigue and P. Batail, Phys. Rev. Lett. {\bf 85}, 5420 (2000).
\bibitem{eld1}
J.E.~Eldridge, Y. Lin, H. H. Wang, J. M. Williams and A. M. Kini ,
Phys. Rev. B {\bf 57} 597 (1998).
\bibitem{wosnitza}
H. Elsinger, J. Wosnitza, S. Wanka, J. Hagel, D. Schweitzer and W.
Strunz , Phys. Rev. Lett. {\bf 84} 6098 (2000); J. Wosnitza,
Physica C {\bf 317-318} 98 (1999).
\bibitem{raman}
D. Pedron, G. Visentini, R. Bozio, J. M. Williams and J. A.
Schlueter, Physica C {\bf 276} 1 (1997).
\bibitem{photonself}
E. Foulques, V. G. Ivanov, C. M\'{e}zi\`{e}re and P. Batail, Phys.
Rev. B {\bf 62} R9291 (2000).
\bibitem{pintschovius}
L. Pintschovius, H. Rietschel, T. Sasaki, H. Mori, S. Tanaka, N.
Toyota, M. Lang and F. Steglich, Europhys. Lett. {\bf 37} 627
(1997); N. Toyota, M. Lang, S. Ikeda, T. Kajitani, T. Shimazu, T.
Sasaki and K. Shibata, Synth. Met. {\bf 86} 2009 (1997).
\bibitem{Y123}
R. M. MacFarlane, H.J. Rosen, E.M. Engler, V.Y. Lee and R.D. Jacowicz,
Phys. Rev. B {\bf 38}, 284 (1988).
\bibitem{kini}
A. M. Kini, K. D. Carlson, H. H. Wang, J. A. Schlueter, J. D.
Dudek, S. A. Sirchio, U. Geiser, K. R. Lykke and J. M. Williams,
Physics C {\bf 264} 81 (1996)
\bibitem{schlueter}
J. A. Schlueter, A. M. Kini, B. H. Ward, U. Geiser, H. H. Wang, J.
Montasham, R. W. Winter, G. L. Gard, Physica C {\bf 351} 261
(2001).
\bibitem{watanabe}
Y. Watanabe, T. Shimazu, T. Sasaki, N. Toyota, Synth. Met. {\bf
86} 1917 (1997).
\bibitem{perscom}
J.A. Schlueter, U. Geiser, unpublished data (2002).
\bibitem{rachel} R.S. Edwards {\it et al.}, submitted
to Phys. Rev. Lett.
\bibitem{eremets}
M.~Eremets, {\it High Pressure Experimental Methods}, (Oxford University Press,
Oxford, 1996).
\bibitem{goddard}
John Singleton, P.A. Goddard, A. Ardavan,
N.~Harrison, S.J. Blundell, J.A.~Schlueter and A.M.~Kini,
Phys. Rev. Lett. {\bf 88}, 037001 (2002).
\bibitem{neilbd}
N.~Harrison, J. Caulfield, J. Singleton, P.H.P. Reinders, F. Herlach,
W. Hayes, M. Kurmoo and P. Day,
{\it J. Phys.: Condens. Matter} {\bf 8} 5415 (1996).
\bibitem{caulfield}
J.M.~Caulfield, W.~Lubczynski, F.L.~Pratt,
J.~Singleton, D.Y.K.~Ko,
W.~Hayes, M.~Kurmoo and P.~Day,
J. Phys.: Condens. Matter {\bf 6}, 2911 (1994).
\bibitem{weiss}
H.~Weiss, M.V.~Kartsovnik, W.~Biberacher,
E.~Steep, A.G.M.~Janssen and N.D.~Kushch,
JETP Lett. {\bf 66}, 202 (1997);
H. Weiss, M.V. Kartsovnik, W. Biberacher, E. Steep, E. Balthes, A.G.M. Jansen,
K. Andres and N.D. Kushch, Phys. Rev. B {\bf 59}, 12370 (1999).
\bibitem{brooks}
J.S. Brooks, X. Chen, S.J. Klepper, S. Valfells, G.J. Athas,
Y. Tanaka, T. Kinoshita, N. Kinoshita,
M. Tokumoto, H. Anzai and C.C. Agosta, Phys. Rev. B {\bf 52}, 14457 (1995).
\bibitem{comment}
Recent fits of this model to de Haas-van Alphen
data at ambient pressure have produced the
values $t_{\bf b}=15.6$~meV, $t_{{\bf c}1}=24.2$~meV and
$t_{{\bf c}2}=20.3$~meV~\cite{goddard}.
Note that these $t$ are
{\it effective} transfer integrals, as opposed
to the bare, unrenormalised transfer integrals used in a
bandstructure calculation; instead,
as they are based on parameters
from de Haas-van Alphen data they will include the
effects of electron-electron and electron-phonon interactions~\cite{eva}.
\bibitem{eva}
N. Harrison, E. Rzepniewski, J. Singleton, P.J. Gee,
M.M. Honold, P. Day and M. Kurmoo, {\it J. Phys.: Condens. Matter}
{\bf 11}, 7227 (1999).
\bibitem{pedron2}
D. Pedron, R. Bozio, J.A. Schlueter, M.E. Kelly, A. M. Kini, J. M.
Williams, Synth. Met. {\bf 103} 2220 (1999).
\bibitem{rahal}
D. Chasseau, J. Gaultier, M. Rahal, L. Ducasse, M. Kurmoo, P. Day,
Synth. Met. {\bf 41} 2039 (1991); M. Rahal, D. Chasseau, J.
Gaultier, L. Ducasse, M. Kurmoo, P. Day, Acta Cryst. {\bf B53} 159
(1997).
\bibitem{lang}
J. M\"{u}ller, M. Lang, F. Steglich, J. A. Schlueter, A. M. Kini,
U. Geiser, J. Mohtasham, R. W. Winter, G. L. Gard, T. Sasaki, N.
Toyota, Phys. Rev. B {\bf 61} 11739 (2000).
\bibitem{sadewasser}
S. Sadewasser, C. Looney, J. S. Schilling, J. A. Schlueter, J. M.
Williams, P. G. Nixon, R. W. Winter, G. L. Gard, Sol. St. Commun.
{\bf 104} 571 (1997).
\bibitem{langmoan}
Reference~\cite{lang} does not distinguish between d8 and h8
samples. The pressure dependence of $T_{\rm c}$ for the d8 salt
extracted from thermal expansion data~\cite{lang} does not
agree with the current direct measurement.


\end{thebibliography}
\end{document}